\documentclass[showpacs,aps]{revtex4}
\usepackage{amssymb}
\usepackage{graphicx,verbatim}
\usepackage{epstopdf}
\usepackage{slashed}
\begin{document}

\title{Entropy, confinement, and chiral symmetry breaking}
\author{John M. Cornwall\footnote{Email cornwall@physics.ucla.edu}}
\affiliation{Department  of Physics and Astronomy, University of California,
Los Angeles CA 90095}

\begin{abstract}
 This paper studies the way in which confinement leads to chiral symmetry
breaking (CSB) through a gap equation. We argue that entropic effects
cut off infrared singularities in the standard confining effective propagator
$1/p^4$, which should be replaced by $1/(p^2+m^2)^2$  for a  finite mass
 $m\sim K_F/M(0)$ [$M(0)$ is the zero-momentum value of the 
running quark mass]. Extension of an old calculation of the author
yields a specific estimate for $m$. This cutoff propagator  shows 
semi-quantitatively  two critical properties of confinement:  1)  a 
negative contribution to  the confining potential coming from  entropic 
forces; 2) an infrared  cutoff required by gauge invariance and CSB itself.
Entropic effects lead to a  proliferation of pion branches and a $\bar{q}q$ 
condensate, and contribute  a negative term $\sim -K_F/M(0)$  to the 
effective pion Hamiltonian allowing for a massless pion in the presence 
of positive kinetic energy and string energy. The resulting gap equation
leads to a well-behaved running constituent quark mass $M(p^2)$ with $M^2(0)
\approx K_F/\pi$.   We include one-gluon terms to get the correct 
renormalization-group ultraviolet behavior, with the improvement that 
the prefactor (related to $\langle \bar{q}q\rangle$) can be calculated 
from the confining solution.  We  discuss an integrability condition 
that guarantees the absence of IR singularities at $m=0$ in Minkowski 
space through use of a principal-part propagator.

  \end{abstract}
\maketitle 

\section{Introduction}

\subsection{General}

The purpose of this paper is to explore possibilities of describing quark chiral symmetry breaking (CSB) with confining forces, in the simplest case of zero temperature and density, the only case we consider here (for finite temperature and density, see \cite{gloz} and references therein), without running into difficult infrared (IR) singularities. CSB has two essential manifestations:  A running quark mass $M(p^2)$ with finite $M(0)$, and a massless pion.   Any such description of CSB with confinement alone must also resolve the dilemma that confining forces (such as the usual linearly-rising potential $K_Fr$, for string tension $K_F$) plus kinetic energy terms apparently have no negative terms that could lead to a massless pion.   

   Many papers have been written on CSB with confinement, but the present approach differs from those known to the author.     There are a number of papers \cite{corn070,addavis,yopr,nekroc,sugasa,prosp,bicnef} that make some attempt to model area-law confinement as it might arise in QCD (as opposed to purely phenomenological effective propagators, NJL models, and so on); all of them make one approximation or another, and ours is no exception.  This paper differs from most of the cited papers by attempting to maintain covariance   and avoiding the use of special gauges, such as Coulomb gauge, within the context of a  Euclidean gap  equation with confining forces.

The present paper makes two major points.  The first is to argue that entropic effects (embodied in large spacetime fluctuations in  worldlines of pions that are composites of massless quarks) may well  be a major source of negative terms necessary for a massless pion with area-law confinement (see \cite{corn070} for an early description of entropic contributions).  These entropic effects come from the  masslessness of the quarks and of the (Goldstone) pion, with the consequence that a pionic $q\bar{q}$ Wilson loop with a large longitudinal separation between initial and final points is highly ramified or branched. Anywhere along the perimeter, large transverse separation of $q$ from $\bar{q}$ is exponentially disfavored because of the consequent large area-law action penalty; separation of more than about $M(0)^{-1}$ practically never occurs.  A linearly-rising potential is irrelevant for larger separations.  Because the long, twisting, and narrow branches are made of massless constituents having a massless bound state, their action per unit length is too small to overcome entropy.  These branches signal the formation of a $\langle \bar{q}q \rangle$ condensate.  [Formation of the condensate as branches of a $q\bar{q}$ Wilson loop may have something to do with the light-cone interpretation \cite{brods} of condensates as objects localized with respect to hadrons; we take no position on this possibility.]  On general grounds we show that the entropic effects should contribute a term $\sim -K_F/M(0)$ to the pion mass, and that if only kinetic energy and linearly-rising potential terms are kept, masslessness of the pion is assured if $M(0)$ has a specific value $\sim K_F^{1/2}$.  That this entropic term is negative is crucial, since other negative terms, such as one-gluon exchange or hyperfine structure, may not be large enough to give the pion a zero mass (or in  other words, to yield CSB).  [We have nothing new to say about the rho meson, whose mass we attribute in the standard way to chromomagnetic hyperfine splitting.] 

The second point is that we can find  a non-singular Euclidean gap equation for the quark running mass $M(p^2)$, an extension of the Johnson-Baker-Willey (JBW) equation \cite{jbw} to confining forces, that exemplifies both negative entropic contributions and an effective IR cutoff, coming from the bound on $q\bar{q}$ separation.    If we use the standard  effective propagator $8\pi K_F\delta_{\mu\nu}/k^4$, representing the propagator  of a fictitious Abelian gluon, well-known IR singularities arise in a standard gap equation.  This is because such a gap equation, with an open quark line,  is not invariant with respect to Abelian gauge transformations (having nothing to do with color) of this fictitious gluon.   We point out that, with the aid of a fictitious heavy ``quark" called $\chi$, with mass $M^2_{\chi}\gg K_F$, it is possible to construct a singularity-free Abelian  gauge-invariant dynamics for quark CSB.  This is done by using a regulated effective propagator $8\pi K_F\delta_{\mu\nu}/(k^2+m^2)^2$   in the Green's function   $G_{\chi q} =\langle|T[\bar{\chi}q(x)\bar{q}\chi(0)]|\rangle$.  Because $G_{\chi q}$ has a closed fermion loop it has the Abelian gauge invariance that leads to a cancellation of the IR singularities \cite{corn053,corn070}, so the $m=0$ limit exists, at least as long as there is a mass gap, that is, that there is CSB for the quark.  In such a case gauge invariance gives a natural IR cutoff in the dynamics at a momentum scale $\sim M(0)$.

 Part of this cancellation of singularities was given long ago \cite{corn070}  in another closed-loop process,   the pion Bethe-Salpeter equation.  There it was shown that the limit $m\rightarrow 0$ in the sum $2M+V(r)$ of the on-shell running quark mass and the static potential based on the regulated effective propagator   exists and is finite; the potential is just $K_Fr$. However, any finite terms surviving at $m=0$ were not investigated.  In the present work we show that there is such a   finite and negative term in the mass  $2M$ of order $ -K_F/M(0)$ that we now identify with an entropic contribution.   Such a negative term is essential for producing a zero-mass pion.   In the mass-regulated potential considered by itself there is, at $r=0$, a negative term $-K_F/m$.  We see that in the $m=0$ limit, the regulator mass reappears as a term that we can interpret as entropic for a {\em finite} value of $m\approx M(0)$.  Moreover, finite $m$ leads to an IR cutoff of the type we expect, which means that there is no reason to insist on a linearly-rising potential much beyond distances $\sim M(0)^{-1}$.  The result is a non-singular description of CSB using the mass-regulated potential with finite $m$, that yields $M^2(0)\approx K_F/\pi$.    With only confining forces, the UV asymptotic behavior is $M(p^2)\sim 1/p^4$, but we    add one-gluon effects  to get the known \cite{lane} renormalization-group (RG) behavior, with the additional feature that we can calculate the prefactor from the confining effects.

As a subsidiary point, we investigate the contributions to CSB from a one-dressed-gluon JBW equation.   If, as previously argued from this equation \cite{corn137}, gluonic effects are not enough for CSB, they are still important because they enhance the zero-momentum quark mass $M(0)$ and give the dominant effects in the ultraviolet (UV).

Finally, we briefly mention another approach to removing the singularities of confining forces, by using a principal-part propagator in Minkowski space.

\subsection{The effective confining propagator and the gap equation}

The effective Euclidean propagator (having nothing to do with the true QCD gluon propagator!) that we will use is:
\begin{equation}
\label{expreg}
D_{eff}(k)_{\mu\nu}\equiv \delta_{\mu\nu}D_{eff}(k);\quad D_{eff}(k)=\frac{8\pi K_F}{(k^2+m^2)^2}.
\end{equation}
with a finite value of $m$.
The $m=0$ limit, or $8\pi K_F/k^4$, of this effective propagator mimics an area law when used in  a world-line action:
\begin{equation}
\label{confact}
\oint\!\mathrm{d}x_{\mu}\,\oint\!\mathrm{d}y_{\nu}\delta_{\mu\nu}\widehat{D}_{eff}(x-y;m=0)
\end{equation}
where the integrals are over closed world lines and $\widehat{D}$ indicates a Fourier transform.   It is precisely an area law in $d=2$, where the (massless gluon) propagator is $\sim\delta_{\mu\nu}/k^2$, and it yields  approximately an area law in $d=4$. (For a discussion of how this approximates an area law in $d=4$ see \cite{corn131}.)  It is gauge-invariant in the sense that if derivative terms are added to the propagator they contribute nothing to the action, because of the integration over closed loops.  For finite $m$ this gauge-invariant action does, of course, depend on $m$, but for infinitesimal $m$ a small change is an Abelian gauge transformation \cite{corn070} and this closed-loop action is both independent of $m$ and non-singular at $m=0$.

This choice of massive effective propagator is not unique; as an example, one might try the two-mass propagator:
\begin{equation}
\label{2mass}
D_{eff}(k)_{\mu\nu}\equiv \delta_{\mu\nu}D_{eff}(k);\quad D_{eff}(k)=\frac{8\pi K_F}{(k^2+m_1^2)(k^2+m_2^2)}.
\end{equation}
If both $m_1,m_2$ are close to $M(0)$ there is little difference from Eq.~(\ref{expreg}).  But if either is very small or very large compared to $M(0)$, either unphysical features appear or CSB is absent.  For example, at $m_1=0$ the two-mass propagator is effectively an attractive Coulomb potential with a very strong coupling that has no physical basis.  So we use only the propagator of Eq.~(\ref{expreg}) in what follows.

The  Euclidean JBW equation we use to describe confining effects is:
\begin{equation}
\label{naiveconf}
M(p^2)=\frac{1}{(2\pi )^4}\int\!\mathrm{d}^4k\,D_{eff}(p-k)\frac{4M(k^2)}
{k^2+M^2(k^2)}
\end{equation}
where $D_{eff}(p-k)$ is the effective propagator of Eq.~(\ref{expreg}).   

It is not hard to see that, just as for the  gluonic gap equation \cite{delscad},  the confining gap equation for  $M(p^2)$ is closely related  to the gauge-invariant Bethe-Salpeter equation for the pion, with confining forces.  Recall that the usual derivation of the  JBW equation, for gluonic exchange rather than confinement, comes from the Schwinger-Dyson equation, with zero bare (current) mass, for the CSB mass term $M(p^2)$ in the quark propagator.  This equation is coupled to the equation for the coefficient of $\slashed{p}$ in this propagator and to various vertices related to the propagator by Ward identities, and in its full glory the coupled equations are quite complex (see, for example, a very recent study of the full Schwinger-Dyson equation for the Landau-gauge quark propagator \cite{papag}).  Extensive use of these Ward identities  and other tools show \cite{delscad} that satisfaction of the full quark self-energy Schwinger-Dyson equation implies the existence of a zero-mass pion, whose wave function is closely related to  the mass function $M(p^2)$.  The confining JBW equation we use is similar, but with gluonic lines replaced by the effective propagator of Eq.~(\ref{expreg}).  For simplicity, and because further accuracy is not warranted,  we consider here (as many authors do) only a simplified form of the self-energy equation in which the coefficient of $\slashed{p}$ is unity.   As with any effective propagator, we only use it at one-loop level.

\subsection{Gluon exchange in the JBW equation}

 In addition to these studies of confinement we briefly explore the usual Euclidean JBW equation for one-gluon exchange, but with a massive gluon \cite{corn137}.  One reason to do this is to set the stage for techniques used in the confining JBW equation.  Another is that within the general framework of the gap equation we use for confinement, but with a one-gluon propagator and running charge rather than the effective confining propagator, it appears that  one-dressed-gluon exchange does not yield CSB for quarks \cite{corn137}, but would do so for adjoint fermions (coupled more strongly by a factor of 9/4 in QCD), as shown in lattice simulations \cite{karsch2}.     This one-gluon result is based on the non-perturbative generation of a dynamical gluon mass \cite{corn076,cornbinpap} $m_g$ of about $2\Lambda$ where $\Lambda \approx$ 300 MeV is the QCD mass scale.  The main effect of the dynamical mass is to reduce the zero-momentum strong coupling $\alpha_s(0)$ to about 0.4-0.5 (with no quarks).   The conclusion that one-gluon CSB does not give CSB is far from unasssailable, because of approximations in the gap equation itself and possible inaccuracies in the claimed dynamical mass and running charge.  Indeed, a preprint that came out as this paper was being written up \cite{papag}, based on an extensive and complicated study of the gap equation in Landau gauge and using Landau-gauge form factors from lattice simulations as input, claims that inclusion of rather subtle ghost effects does lead to gluonic CSB without confinement. As these authors acknowledge, the CSB mechanism they find is not very strong, and they also confirm that CSB is absent within the approximations of Ref.~\cite{corn137}, which has no explicit quark-gluon vertex corrections or ghost contributions.  

In our opinion the fate of gluonic CSB remains to be determined.  There are lattice simulations \cite{deforc} claiming that in $SU(2)$ lattice gauge theory, confinement by center vortices is both necessary and sufficient for CSB, suggesting that standard gluonic effects are not the CSB driving mechanism.  There are also simulations (for example, \cite{cheng}) showing that the CSB phase transition temperature is quite close (not necessarily identical) to the deconfinement phase transition temperature.    Another recent preprint   \cite{bow} muddies the waters further, claiming that in $SU(3)$ lattice-gauge simulations removal of center vortices removes confinement but not CSB, contradicting the $SU(2)$ simulation results of \cite{deforc}.  We do not know how this puzzle will be resolved.  In order to be consistent we continue to use the one-gluon gap equation of Ref.~\cite{corn137}, and add it to the confining gap equation to recover known \cite{lane,politzer}
renormalization group (RG) and operator-product results in the deep UV,  which are that $M(p^2)\sim const. [\ln p^2]^{a-1}/p^2$ (here $a$ is the Lane \cite{lane} constant) for a constant undetermined by the RG or OPE that is proportional to $\langle \bar{q}q\rangle$.  In our combination of confinement and one-gluon effects we can estimate the unknown constant.

\subsection{Another path:  The integrability condition in Minkowski space}

There is another way of treating $m$ as a regulator, at least in Minkowski space. This   is to define the effective propagator analogous to Eq.~(\ref{expreg}) as a principal-part propagator, which has two important effects.     The first is that it makes the quark action of Eq.~(\ref{confact})  real for any $m$; there is no good physical interpretation of an imaginary part of this action.   The second effect is that use of a principal-part effective propagator leads to an integrability condition that allows for the limit $m=0$.   We save a detailed investigation of this propagator, and its extension to Euclidean space, for a later paper.   However, it is important to point out that the principal-part effective propagator obeys a simple condition, called the integrability condition, which is that   the spacetime integral of the confining effective propagator
\begin{equation}
\label{1stintcond}
  \int\!\mathrm{d}^4k\,D_{eff}(p-k)
\end{equation}
is a finite dimensionless constant independent of any regulation scheme used to define the effective propagator.     This real propagator is half the sum of the advanced and the retarded propagators, or the Wheeler-Feynman propagator.    

Our major claim is that whether one uses technique allowing $m$ to be thought of as a true regulator, or whether one identifies a $-K_F/m$ term as a finite entropic contribution in the sort of gap equation we consider here, proper choice of $m$ will lead to correct CSB predictions from such a gap equation.

\subsection{Organization of the paper}

 Section \ref{gisec} shows how to extract the   JBW equation from the Green's function $G_{\chi q}$.  Section \ref{lqarealaw}    develops the entropic argument, basically that  configurations of a Wilson loop with massless quarks heavily favor  highly-ramified structures with pion branches, rather than configurations where the $q$ and $\bar{q}$ can separate by large distances (which would be allowed, in principle, in the quenched approximation where confining strings cannot break). 
Section \ref{masslessp} calculates a previously-neglected regulator-independent term in an old argument \cite{corn070} concerning the cancellation between a regulator in a $1/k^4$ potential and this regulator as it appears in the on-shell quark mass $M\equiv M(p^2=M^2)$ calculated from the one-loop Feynman graph with this potential.  This term, which is negative and $\sim -K_F/M$, is crucial to find a massless pion given that naive confinement contains only the positive area-law potential and kinetic energy; we identify it with entropic contributions. (Negative terms such as one-gluon exchange and hyperfine splitting are too small, or else they would give CSB without confinement.)          Sections \ref{1gjbw},   \ref{reggap} and \ref{1gconf} discuss Euclidean JBW gap equations for, respectively, massive one-gluon exchange; a confining potential with finite regulator mass $m\sim K_F^{1/2}$;   and a JBW equation    for both confining forces and for one-gluon exchange.   Section \ref{arealaw} introduces the integrability condition for removing singularities of a confining effective propagator in Minkowski space. 

\section{\label{gisec} Abelian gauge invariance, infrared cutoffs, and entropy  }

\begin{figure}
\includegraphics[width=4in]{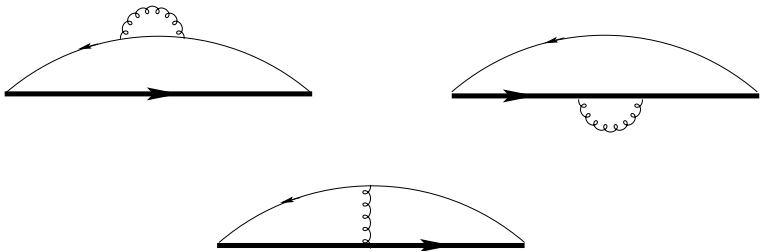}
\caption{\label{gigraphs} Two-loop graphs whose sum has Abelian gauge invariance for the confining propagator.  The thick line is the $\chi$ field; the thin line is a quark with zero current mass; the curly line is the ``area-law" propagator.}
\end{figure}

     We argue that it is possible to extract a JBW gap equation from an Abelian gauge-invariant amplitude that is non-singular and that has all the physical effects (entropy, IR cutoff) represented with reasonable quantitative accuracy.  This extraction is in somewhat the spirit of the pinch technique \cite{corn076,cornbinpap},  in which (color-) gauge-invariant Green's functions are extracted from some gauge-invariant object such as the S-matrix.  Consider the gauge-invariant heavy-light Green's function $G_{\chi q}=\langle |T(\bar{\chi}q(0)\bar{q}\chi(x))|\rangle$, as shown in Fig.~\ref{gigraphs}.    
 As appropriate for an effective propagator, we keep only graphs with a single effective-propagator loop. One can show either by direct calculation of graphs or by using the techniques of \cite{corn053} that such a gauge-invariant color-singlet amplitude  is of the form:
\begin{equation}
\label{corntik}
\int\!\mathrm{d}^4k\,\frac{T_{\mu\mu}(k,\dots)}{(k^2+m^2)^2}.
\end{equation}
Whatever the value of $m$, $T_{\mu\nu}$ is gauge invariant and conserved:  $k_{\mu}T_ {\mu\nu}=0$.  A mass gap plus conservation imply that  $T_{\mu\nu}$ must vanish at $k=0$, and so the $m=0$ singularities  cancel in the sum of all potentials and masses.  The JBW gap equation comes from dropping all terms referring to the field $\chi$ in the sum of graphs, after cancellation of gauge artifacts.

Note that in practice if one uses the gauge-invariant description of CSB through the Green's function $G_{\chi q}$ little is lost by keeping a finite $m$, provided that $m<K_F^{1/2}$, simply because smaller momenta are cut off.  This is exactly what we do in finding the JBW gap equation, because a finite $m\sim M(0)$ incorporates both the gauge-invariance IR cutoff and the entropic term.

We see these properties in the static potential of this effective propagator, which  is:
\begin{equation}
\label{newpotm}  
V(r)=\frac{-K_Fe^{-mr}}{m} = \frac{-K_F}{m}+\frac{K_F}{m}(1-e^{-mr})
\end{equation}
where the first term on the right of the last term is $V(r=0)$.  We will identify this negative term with entropic effects, choosing a specific finite value of $m$.  Note that  the small-$mr$ limit of the remaining finite term is the usual linearly-rising potential $K_Fr$.
The $r$ dependence  at distances $\geq \sim  1/M(0)$, which in this case is virtually flat,   should not matter much because separation of the $q$ and the $\bar{q}$ beyond this distance is very improbable.

\section{\label{lqarealaw} Entropy and a $\bar{q}q$ condensate}

It is rather easy to understand the basic properties of area-law dynamics for {\em heavy} quarks $\chi$, where heavy means that the quark mass $M_{\chi}$ obeys $M_{\chi}^2\gg K_F$. (We assume these quarks are quenched and that there are no other matter fields.)   Such quarks move in essentially classical paths, nearly straight lines, even for times $T$ that can be large compared to the QCD time scale, so that the spatial separation $R$ of a $\chi \bar{\chi}$ pair can be specified in advance with small changes coming from the quark dynamics.  An area law simply means that the   expectation value of the $R\times T$ Wilson loop describing this configuration  is
about $\exp [-K_FRT]$.  There is no question of a condensate of these quarks, since their paths (in Minkowski space) have essentially no backward-moving segments, necessary for a $\langle \bar{\chi}\chi \rangle$ condensate.  Moreover, the entropy of the flux sheets confining these quarks is not large compared to the action (or else the 't Hooft criterion would be violated, yielding both confinement and dual confinement).

For {\em light} quarks $q$, with (current) mass obeying $M^2_q\ll K_F$,   things can be very different.  (Again assume the quenched case, with no other matter fields; string breaking is impossible.)  Suppose we fix the time $T$ between the initial and final $q\bar{q}$ configurations, each for simplicity taken to be at the spatial origin.  What kind of paths occur in the path integral for a quantity such as $\langle T[\bar{q}\gamma_5q(0)\bar{q}\gamma_5q(T,\vec{0})]\rangle$ , and what kinds of areas do these paths have?  

\begin{figure}
\includegraphics[width=4in]{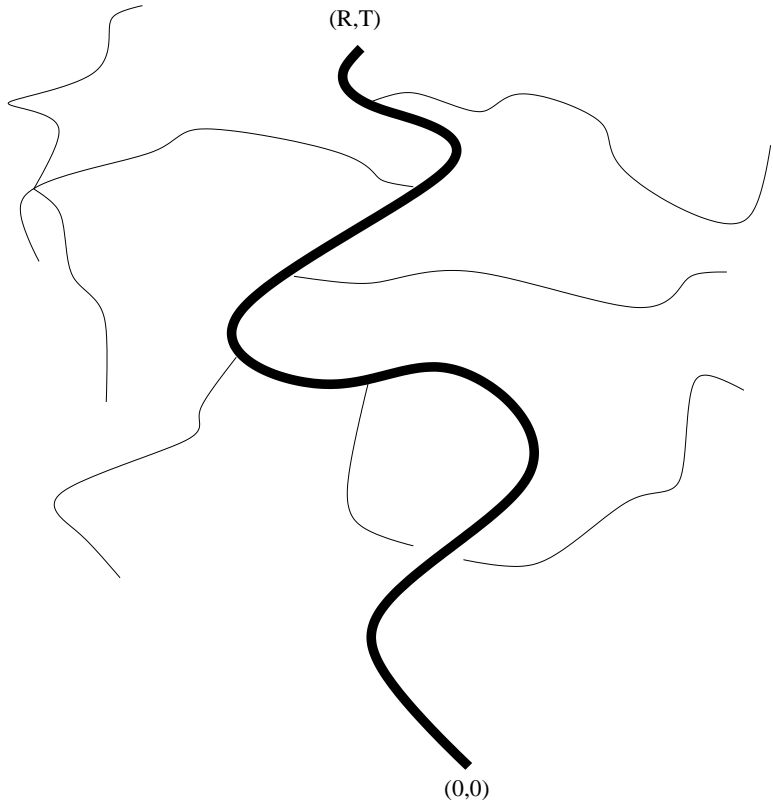}
\caption{\label{branched}  A schematic of a Wilson loop for a pion characteristic of light-quark dynamics; the thick line symbolizes the original $q\bar{q}$ loop, and the thin lines symbolize pions.  Because the loops are narrow we do not show their individual $q$ and $\bar{q}$ lines. A $\gamma_5$ is understood at each end point and every three-vertex.}
\end{figure}

\begin{figure}
\includegraphics[width=4in]{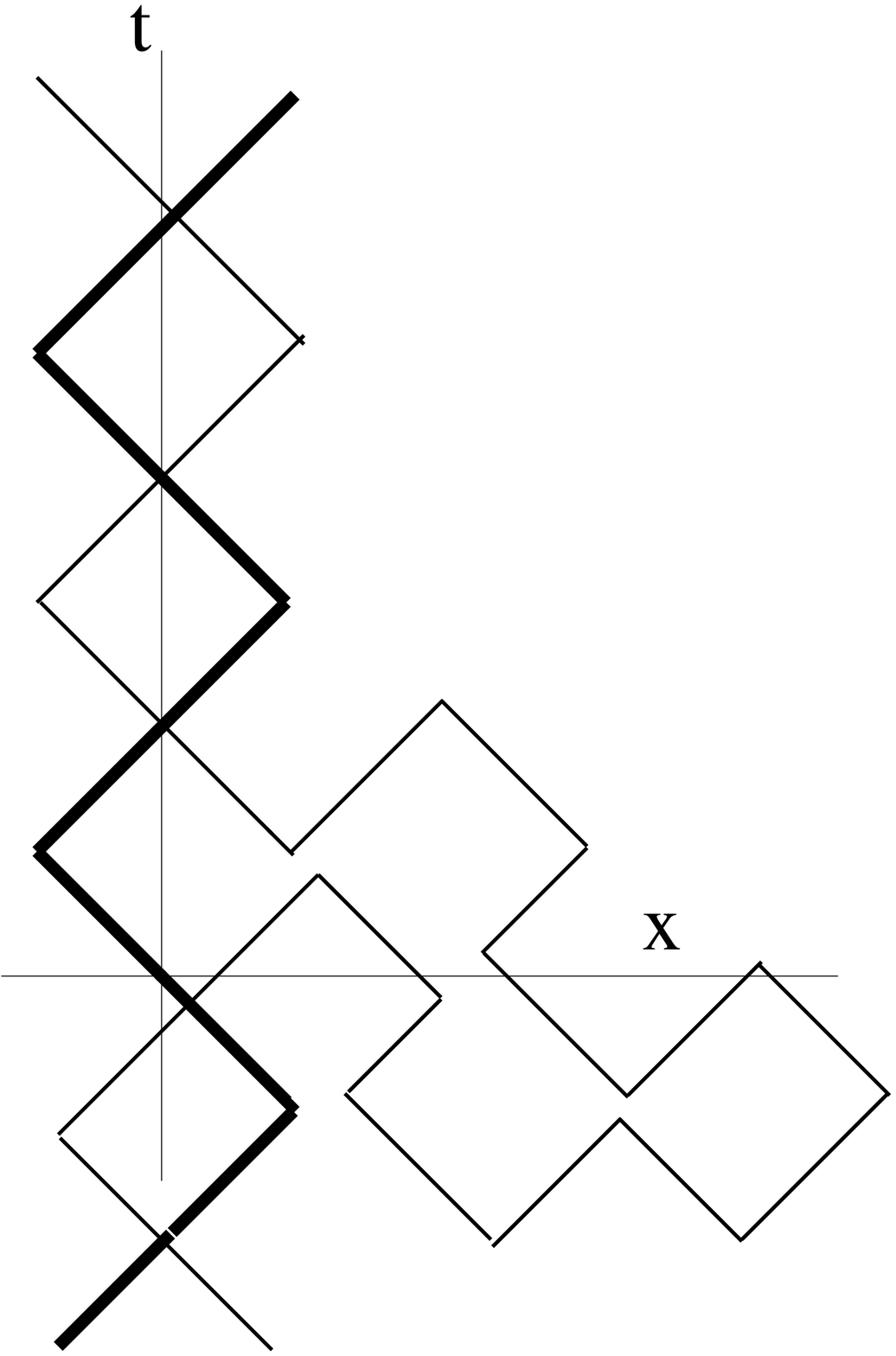}
\caption{\label{entrfig} A schematic closeup of the $q\bar{q}$ Wilson loop in Minkowski space, showing the propagation of massless particles along their light cones under the influence of a confining force (after \cite{corn070}).  Very massive quarks would only propagate in the forward timelike direction, but light quarks can have spacelike legs, signifying formation of a condensate.}
\end{figure}

The answer is that the paths are highly ramified (branched), as shown in Fig.~\ref{branched}.  (For clarity we do not show the fine-scale movement of the quarks about each other, which is sketched in Fig.~\ref{entrfig}.) We can no longer specify the average spatial separation $R$ of these light quarks, which must be calculated.   
An average Wilson loop resembles a highly-branched polymer, with most of the branches representing pions with mass proportional to $M_q$.  Note that going backwards in time is not at all hindered, so that there is no bar to forming a non-zero value of $\langle\bar{q}q\rangle$.  A typical configuration will have an overall   length 
$L\sim T^2/\ell$ , but any particular branch will have a separation between the $q$ line and the $\bar{q}$ line of $\mathcal{O}(\ell )$.  Here the correlation length $\ell$ is related to the CSB-generated constituent quark mass $M$ by $K_F\ell \simeq M$ (see, for example, an elementary discussion in \cite{corn132}).  So the area of the Wilson loop is $\mathcal{O}(L\ell)$.  In terms of the overall length $L$, this looks like a perimeter law, although the actual area can be large because of ramification.      

The upshot of this discussion is that when $K_F\gg M_q^2$, pionic Wilson-loop configurations that look sheet-like (that is,  for some $q\bar{q}$ separation $R$ that is large compared to $K_F^{-1/2}$) are highly suppressed.  This is because the available configurational entropy of the flux sheets cannot overcome the action penalty in the exponentially-small area law contribution to the Wilson loop VEV.   But because the action penalty for forming pions from light quarks is small, entropy can dominate.  Roughly speaking, the entropic term to be added  to the area-law term gives a result of the form:
\begin{equation}
\label{entropy}
\langle W\rangle \sim \exp [-K_FT^2+\frac{T^2}{\ell^2}\ln (2d-1)]
\end{equation}
in spacetime dimension $d$.  Since the overall length $L$ scales as $T^2/\ell$ this can also be written:
\begin{equation}
\label{altentropy}
\langle W\rangle \sim \exp [-K_FL\ell +\frac{L}{\ell}\ln (2d-1)].
\end{equation}
(We use a standard approximation for the entropy, as counting the ways links of length $\ell$ can extend themselves on a hypercubic lattice with no backtracking.  There are other terms in the action coming from quark-pion vertex effects, and other terms in the entropy coming from counting the ways the Wilson loop is ramified; we do not discuss them here.)  With $\ell \sim M/K_F$  the entropic term contributes a term $\sim -K_F/M$ to the action density, or energy.  The Wilson loop will ramify until the two terms are approaching equality, at which point other physical effects take over.    

Even though rho mesons (for example) are made of virtually massless quarks, they are heavy in part because of QCD hyperfine interactions to the point that their fluctuation entropy cannot overcome their action.     A standard estimate of hyperfine splitting is:
\begin{equation}
\label{pirho}
M_{\rho}-M_{\pi} \approx \frac{32\pi \alpha_s(0)}{9M^2}|
\psi (0)|^2
\end{equation}
which is about 700 MeV for parameters that we use in this paper.

\section{\label{masslessp} How can a pion be massless with confining forces?}

For purposes of the following heuristic discussion, we assume that one-gluon effects are too weak to produce CSB by themselves, and omit writing them.   

A long-standing problem of CSB via confinement is that a linearly-rising potential such as $K_Fr$ is positive; when added to the positive kinetic energy of the quarks, how can one get a zero pion mass bound state?    
The answer is the negative entropic contributions of the last section.  We saw there that the effective confining propagator of Eq.~(\ref{expreg}) has such a term, although there was no explicit reference to entropy.

  There is another way to find such a negative term.  The Minkowski-space version of the gap equation, Eq.~ (\ref{naiveconf}), and its IR divergences were studied in \cite{corn070}, but without mentioning an important point that we take up here. In the right-hand side  of this Minkowski-space gap equation  set $M(k^2)$ to a constant value $M$, and calculate the integral on-shell, that is, at $p^2=M^2$.  The result is the running mass on-shell, or $M(p^2=M^2)$.  In a gap equation this should be the same as the input, which leads to, as  given in \cite{corn070}:
\begin{equation}
\label{oldlin}
M=\frac{K_F}{2m}+\dots
\end{equation}
where the omitted terms either vanish at $m =0$ or are independent of it. The static potential is just that of Eq.~(\ref{newpotm}), and for small $m$ it looks like:
\begin{equation}
\label{statpotmu}
V(r)=\frac{-K_Fe^{-mr}}{m}= -\frac{K_F}{m}+K_Fr+ \mathcal{O}(m).
\end{equation}
So $2M+V(r)$ is free of infrared divergences in the $m=0$ limit.    What was not explored in this earlier paper was the $m$-independent term of Eq.~(\ref{oldlin}).  We have now calculated this term; adding it to $2M$ one finds:
\begin{equation}
\label{oldlinplus}
2M+V(r)=K_Fr-\frac{3K_F}{\pi M}.
\end{equation}
In effect, although the $-K_F/m$ term in $V(r)$ cancels, it reappears in finite form as $-3K_F/(\pi M)$.
  At first it may seem odd that the mass operator gives a negative term to the sum $2M+V(r)$, but something must do so, or it will be impossible to find a zero-mass pion (in the present crude approximation).  

Look at the relativistic pseudo-Hamiltonian 
\begin{equation}
\label{pseudoh}
H=p+K_Fr-\frac{3K_F}{\pi M}.
\end{equation}
Substitute $p\rightarrow 1/r$ and minimize on $r$ to find a variational approximation 
\begin{equation}
\label{varapprox}
\langle H \rangle = 2K_F^{1/2}-\frac{3K_F}{\pi M}.
\end{equation}
There is a zero-mass bound state when $M=3K_F^{1/2}/(2\pi )$, or about 220 MeV.      Of course the estimate coming from Eq.~(\ref{varapprox}) is only qualitative, and in the real world there are other negative terms to be included, including gluon exchange and hyperfine structure, but these are not the dominant negative contributions.  (If they were, the one-gluon JBW equation would have yielded CSB, but it is perhaps plausible that it does not.)  

If we identify the two negative constants in Eqs.~(\ref{oldlinplus},\ref{statpotmu}), we find that $m$ is close to $M$:
\begin{equation}
\label{2mequal}
m=\frac{\pi M}{3}.
\end{equation}
This is not necessarily an accurate value, so in  the Euclidean confining gap equation using the Euclidean effective propagator of Eq.~(\ref{expreg}) we set  $m=\alpha M$ for a range of $\alpha \approx$ 1.  But first, both to set the stage for a Euclidean phenomenology of confinement CSB and to illustrate the argument that one-gluon exchange may not be enough for quark CSB, we briefly review the arguments of \cite{corn137} concerning one-gluon exchange.

\section{\label{1gjbw} Massive one-gluon exchange}

   There is no doubt that the JBW equation for QCD with a massless gluon is correct and useful for asymptotically-large momenta \cite{lane}, and in fact we will recover these results as an addition to CSB by a purely confining force.   However, it can be questioned whether one-gluon effects in the IR can produce CSB for quarks  \cite{corn137}.

Long ago the author argued that the infrared singularities of QCD, coming from asysmptotic freedom, had to be cured by the generation of a dynamical gluon mass \cite{corn076}.  The pinch technique (PT) and the gauge technique were used to enforce gauge invariance in off-shell Green's functions for non-Abelian gauge theories.  (For a comprehensive treatment of the pinch technique see \cite{cornbinpap}; the gauge technique constructs vertices as functionals of self-energies such that the Ward identities are exactly satisfied. These vertices are approximate, but expected to be asymptotically accurate at small momenta.)    A one-dressed-loop PT approximation showed that  ``wrong-sign" asymptotic freedom problems were cured by generation of a dynamical gluon mass of about 600-700 MeV or so. In recent years such a dynamical mass has been abundantly confirmed by lattice simulations and more sophisticated PT treatments; see \cite{cornbinpap}.     As for the running charge,   \cite{corn076} gives the following approximation for Euclidean (spacelike) momenta: 
\begin{equation}
\label{qcdcharge1}
\bar{g}^2(k^2)=\frac{1}{b\ln [(k^2+4m_g^2)/\Lambda^2]};\quad \alpha_s(0)=\frac{1}{4\pi b 
\ln (4m_g^2/\Lambda^2)}
\end{equation}
where $m_g$ is the dynamical gluon mass, $\Lambda$ the QCD mass scale, and, for gauge group $SU(N)$ with $N_f$ flavors, $b$ is the one-loop coefficient in the beta-function:
\begin{equation}
\label{beta}
b=\frac{11N-2N_f}{48\pi^2}.
\end{equation} 
  (We use this running charge only  only for spacelike momenta, so the singularity for timelike momentum is irrelevant.  There is a more complicated modified form \cite{corn138} that is free of singularities in the timelike regime, and it agrees rather well with the above form for spacelike momenta.) In this paper we use, consistent with lattice determinations, phenomenology, and more sophisticated treatments of the PT Schwinger-Dyson equations \cite{cornbinpap}, $m_g=2\Lambda\approx$ 600 MeV. 

Let us accept these PT results, although the final word has yet to be said on their quantitative accuracy, and ask what they have to say about one-gluon CSB.   Recently the author argued
 \cite{corn137} that the gluon mass and relatively small coupling were, in fact, too small for quark CSB via the gluon-exchange JBW mechanism, and confinement had to be the main source of CSB, as we now review.  
  The {\em linearized}  massive-gluon JBW equation  is,  in Landau gauge:
\begin{equation}
\label{unregint}
M(p^2)=\frac{C_2g^2}{(2\pi )^4}\int\!\mathrm{d}^4k\,\frac{3M(k)}{[(p-k)^2+m_g^2]k^2}
\end{equation}
where $C_2$ is the quark Casimir eigenvalue (4/3 in $SU(3)$) and $m_g$ the gluon mass.      It turns out that accounting for a gluon mass does two things:  1)  It makes the gap equation finite at zero momentum; 2) and more important, it bounds the IR running coupling.  If the mass is too small, the running charge gets unacceptably large, as judged by phenomenology and solutions to the PT Schwinger-Dyson equations.

In the IR we can replace the running charge $\bar{g}^2(q^2)$ by its zero-momentum value, which we call simply $g^2$, and then integrate over angles, with the result:
\begin{equation}
\label{angint}
\int\!\mathrm{d}\Omega_k \frac{1}{(p-k)^2+m_g^2}\equiv K(k;p)=\frac{4\pi^2}{p^2+k^2+m_g^2+[(p^2+k^2+m_g^2)^2-4p^2k^2]^{1/2}}.
\end{equation}
However, this kernel does not yield a simple differential equation.  Earlier, it was proposed \cite{corn137,corn138} to approximate the angular integral by: 
\begin{equation}
\label{kernapprox}
K(k;p)\approx 2\pi^2\bigg[\frac{\theta (p^2-k^2)}{p^2+m_g^2}+\frac{\theta (k^2-p^2)}{k^2+m_g^2}\bigg]\equiv \widetilde{K}(k,p).
\end{equation}
Numerically the approximate kernel $\widetilde{K}$ is, on the average, about 20-30\% larger than the true kernel for IR momenta ($\leq m_g$), but it approaches the true kernel in the UV. We ignore this IR discrepancy, because it is in the direction to reinforce our conclusion that one-gluon exchange is too weak for CSB, and because the primary use of the one-gluon JBW equation will be for large momenta.
Using the approximate kernel $\widetilde{K}$ and the appropriate arguments for the running charge yields an integral equation:
\begin{equation}
\label{jbwint}
M(p^2)=\frac{3C_2g^2}{16\pi^2}\int\!\mathrm{d}k^2\,\frac{k^2 M(k^2)}{k^2+M^2(k^2)} 
\bigg[\frac{\theta (p^2-k^2)}{p^2+m_g^2}+(k \leftrightarrow p)\bigg].
\end{equation}
There is a corresponding differential equation:
\begin{equation}
\label{jbwdiff}
M''(p^2)+\frac{2M'(p^2)}{p^2+m_g^2}+\frac{\lambda M(p^2)}{(p^2+m_g^2)^2}=0;\quad \lambda = \frac{3C_2g^2}
{16\pi^2}.
\end{equation}
This is nothing but the original JBW equation, with the variable $p^2+m_g^2$ in place of $p^2$.  It has power-law solutions
\begin{equation}
\label{powerlaw}
M(p^2)= \textrm{const.}(p^2+m_g^2)^{\nu_{\pm}},\quad \nu_{\pm}=\frac{1}{2}\{-1\pm[1-4\lambda ]^{1/2}].
\end{equation}
If the zero-momentum coupling is too small, there is no CSB.  The standard analysis is that the critical coupling is the point at which the square root in Eq.~(\ref{powerlaw}) turns imaginary, and  that there is CSB for couplings larger than this critical value.  Then one-gluon CSB, in this approximation, requires:
\begin{equation}
\label{critcoup}
\alpha_s(0)\equiv \frac{g^2}{4\pi}\geq \frac{\pi}{3C_2}.
\end{equation}
With $C_2=4/3$ this yields $\alpha_s(0)\geq 0.8$, approximately, somewhat greater than the value 0.5 given by Eq.~(\ref{qcdcharge1}).  Taking account of the difference between the true massive kernel and the approximate kernel would change the critical value of $\alpha)s(0)$ to about one.  It seems likely, then, that one-gluon exchange is not strong enough to drive CSB.

There are two results of this subsection:  First, the suggestion that ordinary one-gluon exchange is too small to drive CSB for quarks (but likely large enough for adjoing fermions, where the critical coupling is only 4/9 as large, because of the Casimir eigenvalue \cite{corn137}).  Second, one can derive a regulated gap equation for an area law by differentiating with respect to $m_g^2$, and replacing the coupling $g^2$ by $-8\pi K_F$.  We will pursue this second course in order to propose another form of an area law potential, partly based on the considerations of Sec.~\ref{arealaw}.

\section{\label{reggap}  The confining gap equation}

The issue is to solve the equation:
\begin{equation}
\label{gapeq}
M(p^2)=\frac{2K_F}{\pi ^3}\int\!\mathrm{d}^4k\, \frac{M(k^2)}{[k^2+m^2]^2
[k^2+M^2(k^2)]}
\end{equation}.

Clearly, we can find the S-wave projection of the gap equation by differentiating the massive kernel of Eq.~(\ref{angint}) with respect to the gluon mass.  But (as with the one-gluon equation) the resulting gap equation can only be studied numerically.  Again without losing any essential accuracy, we will instead  differentiate the approximate massive one-gluon S-wave kernel $\widetilde{K}$ of Eq.~(\ref{kernapprox}) with respect to $m_g^2$, replace the gluon mass $m_g$ by $m$, and make some other obvious changes.  This yields a Euclidean gap equation:
\begin{eqnarray}
\label{newinteq1}
M(p^2) & = & \frac{2K_F}{\pi (p^2+m^2)^2}\int_0^{p^2}\!\mathrm{d}k^2\frac{k^2M(k^2)}{k^2+M^2(k^2)}
  +\frac{2K_F}{\pi}\int_{p^2}^{\infty}\!\mathrm{d}k^2\frac{k^2M(k^2)} {(k^2+m^2)^2(k^2+M^2(k^2))}\\ \nonumber
& = & J_>(p^2)+J_<(p^2)
\end{eqnarray}
where $J_>(p^2)$ is the integral from 0 to $p^2$.
The corresponding differential equation is:
\begin{equation}
\label{newdiffeq1}
M(p^2)''+\frac{3M(p^2)'}{p^2+m^2}+\frac{4K_F}{\pi}\bigg[\frac{p^2M(p^2)}{(p^2+m^2)^3[p^2+M^2(p^2)]}\bigg]=0.
\end{equation}
This is really a family of differential equations, as we see by writing them in non-dimensional form.  Write 
\begin{eqnarray}
\label{nondim}
M(p^2) & = & Mf(u\equiv p^2/M^2),\quad M\equiv M(p^2=0);\\ \nonumber 
m & \equiv & \alpha M
\end{eqnarray}
to find:
\begin{equation}
\label{diffamily}
f''(u)+\frac{3f'(u)}{u+\alpha^2}+\frac{4K_F}{\pi M^2}\bigg[\frac{uf(u)}{(u+\alpha^2 )^3[u+f^2(u)]}\bigg]=0.
\end{equation}
The boundary conditions are $f(0)=1,f'(0)=0$.  From Eq.~(\ref{2mequal}) we expect $\alpha \approx 1$. Suppose for the sake of argument that $\alpha$ is fixed even if $M$ changes; this is reasonable, given that $m$ is not an externally-imposed quantity but one which actually should scale with $M$.  Then note that the dimensionless coupling parameter $K_F/M^2$ appearing in the differential equation is not known if only the differential equation is available.  It has to be determined from the integral equation at zero momentum:
\begin{equation}
\label{massint}
M^2=\frac{2K_F}{\pi}\int_0^{\infty}\!\mathrm{d}u\frac{uf(u)}{(u+\alpha^2 )^2[u+f^2(u)]}\equiv \frac{
2K_F}{\pi}I(\alpha).
\end{equation}

Either the integral or the differential equation yields a large-momentum falloff $M(p^2)\sim 1/p^4$.  That the UV falloff is faster than  what is expected from the OPE and the renormalization group  is not a problem; the  required behavior $M(p^2)\sim 1((\ln p^2)^a/p^2$  \cite{lane} follows from one-gluon exchange, which we take up in the next section.      While there is no need for  an area law to give the actual UV behavior, it is necessary that the area-law UV behavior be no slower than prescribed by the OPE and the renormalization group.  The correct RG behavior will be reinstated by adding the one-gluon terms of Eq.~(\ref{unregint}).  

For small momentum one finds:
\begin{equation}
\label{noninc}
 f(u)= \big[1-\frac{2K_F}{3\pi M^2}\big(\frac{u^3}{\alpha^6}\big)+\dots \big]
\end{equation}
showing that the running mass changes but little near zero momentum.

\begin{figure}
\includegraphics[width=4in]{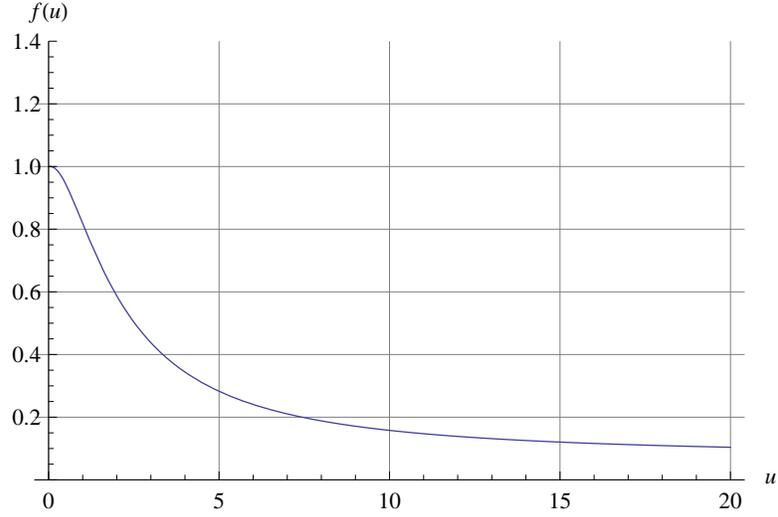}
\caption{\label{areafig}  Numerical calculation of the running quark mass $f(u)$ {\em vs.} $u$, for $M^2=0.8K_f/\pi$ and $\alpha = 0.9$.  }
\end{figure}

 It remains to determine $M$, a quantity that comes entirely from $J_<$.   
Given the family of solutions to the differential equation, we can now estimate the mass by imposing the integral condition of Eq.~(\ref{massint}).  It turns out numerically that $M$ does not change very rapidly with $m$ in the vicinity of $m=M$.  Calculations for the range $0.8<\alpha <1$ yield mass values in the range $M^2=(0.6 -1)K_F/\pi$, with smaller $\alpha$ corresponding to larger $M$.  (The limit $\alpha$ or $m=0$ gives $M$ diverging like $\ln (1/m)$).  We show the case $\alpha$ = 0.9, for which $M\approx 0.9\sqrt{K_F/\pi}\approx 230$ MeV, in Fig.~\ref{areafig}.  Note that $p^2/M^2\approx 15$ corresponds to $p^2\approx$ 1 GeV$^2$.  

As we have seen, at fixed $\alpha$ the only dimensionless strength parameter in this zero-temperature, zero-density problem is $K_F/M^2$.  Unlike the dimensionless coupling constant of one-gluon exchange, this parameter cannot be tuned, since $M$ ultimately is determined by an integral equation.   So there can be no phase transition between CSB and not-CSB at some finite value of this parameter.  If one tries to find a phase with no CSB, so that $M$ somehow approaches zero, the dimensionless strength parameter becomes infinitely strong, and there surely must be CSB.

\section{\label{1gconf}A gap equation with both area law and one-gluon terms}

 One-gluon exchange determines the UV asymptotic behavior through the running charge, so we reinstate $\bar{g}^2$ instead of the zero-momentum value $g^2$ in the one-gluon equation. 
Then the area-law plus one-gluon integral equation is:
\begin{eqnarray}
\label{suminteq}
M(p^2) & = & \int_0^{p^2}\!\mathrm{d}k^2\frac{k^2M(k^2)}{k^2+M^2(k^2)}
\bigg[\frac{2K_F}{\pi (p^2+M^2)^2 }+\frac{3C_2\bar{g}(p^2)^2}{16\pi^2(p^2+m_g^2)}\bigg]\\ \nonumber
 & + & \int_{p^2}^{\infty}\!\mathrm{d}k^2\frac{k^2M(k^2)} { (k^2+M^2(k^2))}   
 \bigg[ \frac{2K_F}{\pi (k^2+M^2)^2}  + \frac{3C_2\bar{g}(k^2)^2}{16\pi^2(k^2+m_g^2)}\bigg] \\ \nonumber
& \equiv & J_>(p^2)+J_<(p^2)+K_>(p^2)+K_<(p^2)
\end{eqnarray}
where $J_>,J_<$ are defined as before (see Eq.~(\ref{newinteq1})) and $K_>,K_<$ refer respectively to the integrals from 0 to $p^2$ and from $p^2$ to infinity of the one-gluon kernel.
Note that for QCD the one-gluon term begins to dominate the confining term at a momentum $p^2$ of order  $2 K_F/\alpha_s(0)$.

One-gluon corrections to confinement are of two types.  The first   correction is in the IR; at zero momentum it comes entirely from $K_<$.   Let us now call the mass coming solely from confining effects $M_c$.  The correction to $M_c$ is approximately:
\begin{equation}
\label{add1g}
 M=M_c\frac{1}{[1-aI_1]};\quad I_1=\int_0^{\infty}\!\mathrm{d}u
\frac{uf(u)}{(u+\gamma)(u+f^2(u))\ln [\beta (u+4\gamma )]}
\end{equation}
where  $\beta = M^2/\Lambda^2,\;\gamma = m_g^2/M^2$, 
and $a$ is the Lane constant:
\begin{equation}
\label{defa}
a=\frac{3C_2}{16\pi^2b}=\frac{9C_2}{11N-2N_f}.
\end{equation}
   As expected, one-gluon corrections increase $M$, since one-gluon effects work in the direction of producing CSB.    For the case shown in Fig.~\ref{areafig}, plus $\beta \approx 0.7,\; \beta \gamma = 4,$ we find $I_1\approx$ 0.22, and with (for no quarks, $SU(3)$) $a$ = 4/11 the mass is increased by a factor of 1.1, approximately, or to about $\sqrt{K_F/\pi}\approx$ 250 MeV.  

The second, and more important to us, is the UV correction, where the one-gluon term dominates.  There is one useful simplification:  The term $K_<$ (last term on the right of Eq.~(\ref{suminteq})) is non-leading by one power of $\ln p^2$ in the UV, and  we will drop it.  Define $M_c(p^2)$ as the solution to the pure confinement equation (\ref{newinteq1}) with a kernel $J_c(p;k)$:
\begin{equation}
\label{mcdef}
M_c(p^2)=\int\!J_c(p;k)M_c(k^2);\quad J_c(p;k)=\frac{2K_Fk^2}{\pi}\bigg[\theta (p^2-k^2)\frac{1}{(p^2
+m^2)^2[k^2+M_c^2(k^2)]^2}+(k\leftrightarrow p)\bigg]
\end{equation}
where
\begin{equation}
\label{defint}
\int \equiv \int_0^{\infty}\!\mathrm{d}k^2.
\end{equation}
Write the solution to Eq.~(\ref{suminteq}), without the $K_<$ term,  as $M(p^2)=M_c(p^2)+Q(p^2)$.  We are only interested in the UV behavior of this equation, which then can be linearized in $Q$, and takes the form:
\begin{equation}
\label{linconf}
M_c(p^2)+Q(p^2)=\int\!J_c(p;k)[M_c(k^2)+Q(k^2)]+\int\!K_>(p;k)[M_c(k^2)+Q(k^2)]
\end{equation}
where the running mass in the denominator of $K_>$ is $M_c(k^2)$:
\begin{equation}
\label{kcdf}
K_>(p;k)= \frac{ak^2\theta (p^2-k^2)}{p^2+m_g^2}\bigg[\ln \big[\frac{p^2+4m_g^2}{\Lambda^2}\big][
k^2+M_c^2(k^2)]\bigg]^{-1}.
\end{equation}
In Eq.~(\ref{linconf}) the $M_c$ on the left cancels against the $J_cM_c$ term on the right, leaving:
\begin{equation}
\label{aftercanc}
Q=J_cQ+K_>(M_c+Q)
\end{equation}
using a streamlined matrix notation, with momentum arguments and integral sign suppressed (this should not lead to confusion of $M_c$ as we now use it with the mass as defined in Eq.~(\ref{defint})).  We will solve this in the UV as a power series in $K_>$.  It will turn out that $Q(p^2)\sim (\ln p^2)^{a-1})/p^2$ in the UV, and it is straightforward to see that the term $J_cQ$ vanishes more rapidly than this, by a power of $p^2$.  So we drop the $J_cQ$ term in Eq.~(\ref{aftercanc}).  This leaves:
\begin{equation}
\label{dropterm}
Q=K_>M_c +K_>Q=\frac{1}{1-K_>}K_>M_c=(1+K_> +(K_>)^2+\dots )K_>M_c.
\end{equation}

Defining the inverse $(1-K_>)^{-1}$ is slightly subtle, because the integral of $K_>$ over a function depends on how rapidly the function vanishes in the UV.  In particular,
the confining solution $M_c$ vanishes like $1/p^4$  in the UV, which means that the function $K_>M_c$ vanishes like $1/p^2\ln p^2$:
\begin{equation}
\label{aftercanc1}
K_>M_c(p^2)=\frac{a}{[p^2+m_g^2 ]\ln \big[\frac{p^2+4m_g^2}{\Lambda^2}\big]}\int_0^{p^2}\!\mathrm{d}k^2\,k^2 M_c(k^2)[k^2+M_c^2(k^2)]^{-1}. 
\end{equation}
Because of the rapid vanishing of $M_C(k^2)$ in the UV we can the upper limit in the integral from $p^2$  to infinity.  Then:
\begin{equation}
\label{upperinf}
K_>M_c(p^2)\equiv \frac{aI_c}{[p^2+m_g^2]\ln \big[\frac{p^2+4m_g^2}{\Lambda^2}\big]}
\end{equation}
 with
\begin{equation}
\label{limtoinfty}
I_c=\int_0^{\infty}\!\mathrm{d}k^2\,k^2 M_c(k^2)[k^2+M_c^2(k^2)]^{-1}.
\end{equation}
  But for any function $F(p^2)$ that behaves like $1/p^2\ln p^2$ at infinity the integral $K_>F$ behaves like $(\ln \ln p^2)/p^2\ln p^2$, and for functions $F$ going like $(\ln \ln p^2)^N/p^2$, $K_>F$  behaves like  $(\ln \ln p^2)^{N+1}/Np^2\ln p^2$.  One can then define the inverse $(1-K_>)^{-1}$ by the pedestrian means of summing the series in Eq.~(\ref{dropterm}), with the result (valid in the UV):
\begin{equation}
\label{lastresult}
Q(p^2)=\frac{aI_c}{p^2\ln p^2}\exp [a \ln \ln p^2]=\frac{aI_c}{p^2\ln p^2}[\ln p^2]^a.
\end{equation}
It is clear that the meaning of $\ln \ln p^2$ in the UV is:
\begin{equation}
\label{logmean}
\ln \ln p^2 \rightarrow \ln [\frac{g^2}{\bar{g^2}(p^2)}]
\end{equation}
in view of:
\begin{equation}
\label{runcharge1}
\frac{g^2}{\bar{g}^2(p^2)}=1+bg^2\ln [1+\frac{p^2}{4m_g^2}]
\end{equation}
which follows from Eqs.~(\ref{qcdcharge1}), (\ref{qcdcharge1}).    Then the result for the UV behavior coming from combining confinement and one-gluon terms is:
\begin{equation}
\label{final}
M(p^2)\rightarrow \frac{3C_2\bar{g}^2(p^2)I_c}{16\pi^2 p^2}\big[\frac{g^2}{\bar{g}^2(p^2)}\big]^a.
\end{equation}
This UV behavior is what the RG dictates, and by adding in confinement effects we are able to give the prefactor $aI_c$.  From the work of \cite{politzer} this allows an estimate of the $\langle \bar{q}q\rangle$ condensate, although we will not pursue that further here because of various complications (see, for example, \cite{chet}).
Finally, we can change $M_c(p^2)$ to $M_{1c}(p^2)$, which is $M_c(p^2)$ modified by   the IR one-gluon corrections ({\em cf.} Eq.~(\ref{add1g}) to incorporate the IR corrections from one-gluon exchange.

\section{\label{arealaw} The integrability condition}

Is it possible to impose an integrability condition on a confining effective propagator, of the form:
\begin{equation}
\label{intcrit}
\frac{1}{(2\pi )^4}\int\!\mathrm{d}^4k\,D_{eff}(p-k)=\textrm{const.}
\end{equation}
where the constant is finite and independent of any regulator masses that were introduced?  If so, the explicit regulator $m$ can be dropped, and the integrability condition provides an implicit regulator that makes the Minkowski-space JBW equation finite.     

The criterion of integrability certainly fails for the Feynman propagator $1/[(p-k)^2-m^2+i\epsilon ]^2$, whether or not there is a regulator $m$.  But it does hold for the principal-part propagator.  The principal part propagator is physically interesting for other reasons, because it is real and leads to no decay processes for the supposed area of the area law.  It is the sum of the advanced and retarded propagators, discussed long ago by Wheeler and Feynman in their preliminary formulation of QED.  

Begin with the standard principal-part propagator known as $\bar{D}(x)$ with mass $m$, defined as:
\begin{equation}
\label{bard}
\bar{D}(x)=\frac{1}{2}[D_{adv}(x)+D_{ret}(x)]=\frac{1}{(2\pi )^4}\int \!\mathrm{d}^4k\,e^{-ik\cdot x}P
\frac{1}{k^2-m^2};
\end{equation}
in space-time it has the value:
\begin{equation}
\label{bardspace}
\bar{D}(x)=\frac{-1}{4\pi}\delta (x^2)+\frac{m\theta (x^2)}{8\pi \sqrt{x^2}}J_1(m\sqrt{x^2}).
\end{equation}
To go to a confining potential, multiply by $8\pi K_F$ and differentiate with respect to $m^2$:
\begin{eqnarray}
\label{dalprop}
D_{eff}(x)  & = & -8\pi K_F\frac{\partial \bar{D}(x)}{\partial m^2}\\ \nonumber
& = & \frac{-8\pi K_F}{(2\pi)^4}\int\!\mathrm{d}^4kP\bigg[\frac{1}{(k^2-m^2)^2}
\bigg] e^{-\mathrm{i}k\cdot x} \\ \nonumber
~ & = & \frac{-K_F}{2}\theta (x^2)J_0(m\sqrt{x^2}).
\end{eqnarray}
The static potential, the time integral of $D_{eff}$, is the usual $K_Fr$ plus an infinite constant coming from the integral over all time.  So we have not entirely exorcised the singularities of a confining potential.  Earlier it was argued \cite{corn070} that this singular term can be gauged away, with a singular gauge transformation.  

The integrability condition will be studied in detail in a later work.  For now it is enough to say that if in the Minkowski-space gap equation we make the drastic approximation $\theta (x^2)\rightarrow 1$ and certain other simplifications, the result is a wrong-sign $\mathcal{M}^4$  theory with a homogeneous solution 
$\mathcal{M}^2=K_F/\pi$.  Here $\mathcal{M}(x)$ is the Fourier transform of $M(p^2)$.

A potential problem shows up:  How to continue such a propagator to Euclidean spacetime for use in a Euclidean JBW equation?  There is no obvious continuation of the theta-function.  If there is a reasonable continuation, will it preserve an integrability  condition?   We have found infinite classes of (the S-wave projections of) Euclidean propagators with no regulator mass that satisfy an integrability condition and behave as $8\pi K_F/k^4$ for certain ranges of momenta $<K_F^{1/2}$, and found that the integrability condition insures singularity-free CSB.   One example:  In the integral equation (\ref{newinteq1}) multiply the integrand of $J_<(p^2)$ by $p^2/k^2$, and then set $m=0$ in both $J_<$ and $J_>$.  The resulting massless kernel yields a finite constant when integrated over $k^2\;\mathrm{d}k^2$, and the corresponding gap equation has properties very similar to those found in connection with the original kernel of Eq.~(\ref{newinteq1}).

\section{Summary}

We suggest that when   chiral symmetry is broken, leading to a running quark mass $M(p^2)$, the ensuing massless Goldstone bosons contribute to an entropy-driven condensate such as $\langle \bar{q}q\rangle$ by ramifying a large number of branches from a basic ``trunk" Wilson loop that itself shows large fluctuations.  Even if the Wilson loop represents quenched quarks, and so is incapable of breaking, configurations with the $q$ and the $\bar{q}$ far apart are quite improbable compared to ramified configurations where they are only separated by a distance of order $K_F^{-1/2}$, because the associated area-law action is large compared to the entropy.  This means that in a linearly-rising potential $K_Fr$, separations with $r\gg K_F^{-1/2}$ are not probed, and the potential at such large distances is irrelevant.  In turn, this means that confinement dynamics need a linearly-rising potential only out to a finite distance $r\sim K_F^{-1/2}$.   

From these considerations we formulate a non-singular confining gap equation of JBW type, with a mass $m$ in the confining effective propagator treated not as a regulator to be set to zero, but as a finite mass $\sim M(0)$ that can be estimated from processes that respect the Abelian gauge invariance associated with the confining effective propagator.  The static potential coming from the confining effective propagator rises linearly only out to a finite distance, and has a negative term at the origin that we identify with entropic effects.  We estimate $m$ through comparison with an extension of an old calculation (having Abelian gauge invariance) in which $m$ could be properly treated as a regulator mass, to be sent to zero after cancellations. The extended calculation replaces the regulator by a specific and physical mass, whose dynamical effects are equivalent to keeping $m$ as finite and of this value in the JBW equation.  

We also studied one-gluon effects within the same framework, thereby finding the correct large-momentum behavior of QCD as known from the RG, but with a calculable prefactor.  The final result, including IR  one-gluon enhancements, is a quark mass $M(0)\approx$ 250 MeV.  This is somewhat smaller than the commonly-quoted value of 300 MeV, which is largely based on not on true dynamical estimates but on assuming that the sum of quark masses represents most of the mass of the hadron in question.  Our mass has a different interpretation, given its entropic underpinnings.

Much remains to be done, both in formulating and solving more elaborate (and more accurate) forms of the gap equation, including in Minkowski space; relating these to pion dynamics; and perhaps making some progress in understanding entropic effects quantitatively.   
\begin{acknowledgments}

A version of this work was presented at the symposium {\em QCD and Strings:  Elements of a Universal Theory}, Oberw\"olz, Austria, September 2010.   I thank the organizers for the opportunity to present this work.  
\end{acknowledgments}

\end{document}